\documentclass[aps,prb,groupedaddress,longbibliography,preprint]{revtex4-1}

\usepackage{graphicx}
\usepackage{dcolumn}
\usepackage{bm}

\begin{document}

\title{Variational approach to the stationary spin-Hall  effect}

\author{J.-E. Wegrowe}
\affiliation{Laboratoire des Solides Irradi\'es, Ecole polytechnique, CNRS, CEA, Universit\'e Paris-Saclay, F 91128 Palaiseau, France}
\author{P.-M. D\'ejardin}
\affiliation{LAMPS, Universit\'e de Perpignan Via Domitia, 52 Avenue Paul Alduy, F 66860 Perpignan, France}


\date{\today}

\begin{abstract}
The Kirchhoff-Helmholtz principle of least heat dissipation is applied in order to derive the stationary state of the spin-Hall effect. Spin-accumulation due to spin-orbit interaction, spin-flip relaxation, and electrostatic interaction due to charge accumulation are treated on an equal footing. A nonlinear differential equation is derived, that describes both surface and bulk currents and spin-dependent chemical potentials. It is shown that if the ratio of the spin-flip relaxation length over the Debye-Fermi length is small, the stationary state is defined by a linear spin-accumulation potential and zero pure spin-current. 
\end{abstract}

\pacs{72.25.Mk, 85.75.-d \hfill}

\maketitle

The classical bulk spin-Hall effect (SHE) is an ohmic conduction process occurring in non-ferromagnetic conductors, in which spin-orbit interaction leads to a spin-accumulation process \cite{Dyakonov,Dyakonov2,Hirsch,Zhang,Tse, Maekawa, Review,Saslow,JPhys,EPL}.  In the framework of the two channel model \cite{Hirsch,Zhang,Tse, Maekawa}, the system can be described as two sub-systems equivalent to two usual Hall devices, with an effective magnetic field that is acting in opposite directions (see Fig.1). In the Hall bar geometry \cite{EPL}, charge accumulation is produced inside each spin-channel over the Debye-Fermi length scale $\lambda_D$. Due to the symmetry of the spin-orbit effective field, the total charge accumulation for the two channels cancels out and the total electric potential between the two edges of the device is zero. However, spin accumulation of both channels adds up and the consequences can be exploited in terms of spin-accumulation  \cite{Awschalom,Jungwirth,Valenzuela,Otani,Gambardella,Bottegoni}. 

In the usual descriptions of SHE \cite{Dyakonov,Dyakonov2,Hirsch,Zhang,Tse, Maekawa, Review,Saslow,JPhys}, the system is defined with two sets of equations: the Dyakonov-Perel (DP) transport equations and the conservation laws for the spin-dependent electric charges. However, as far as we known, the conservation equations used in order to describe the drift currents in both spin-channels and the spin-flip relaxation from one-channel to the other do not take into account the electrostatic interaction and screening effects that govern the electric potential along the $y$ axis. Indeed, the conservation equation used for SHE is that derived in the framework of spin-injection effects \cite{Valet-Fert, PRB2000,Shibata, Zhang,Tse,Maekawa,Review, Saslow,JPhys} - i.e. without electric charge accumulation - that leads to a spin-accumulation spreading over the typical length scale $l_{sf}$. 

In order to derive the equations corresponding to the stationary states of the SHE, we apply the second law of thermodynamics through the Kirshhoff-Helmholz principle of least heat production \cite{Jaynes}.  All three fundamental components of the SHE are taken into account on equal footing. Namely: the effect of the effective magnetic field due to spin- orbit scattering, the electric charge accumulation with electrostatic interactions and screening, and the spin-flip relaxation effect described by the chemical potential difference between the two spin-channels.  A nonlinear fourth order differential equation is then derived for the chemical potentials, that describes non-trivial spin-currents flowing at the surface (defined over the length $\lambda_D$), and the bulk spin-dependent electric fields. In the case of small charge accumulation, it is shown that the stationary state is reached for linear spin-accumulation potential and zero pure spin-current at the limit $l_{sf} \gg \lambda_D$. 

This work shows that the variational approach yields a firm basis for the modeling of complex phenomena occurring in spintronic devices, like spin Hall magnetoresistance, Spin-pumping effect, and Spin-Seebeck or spin-Peltier effects.\\
 \begin{figure} [h!]
   \begin{center} 
   \begin{tabular}{c}
\includegraphics[height=8cm]{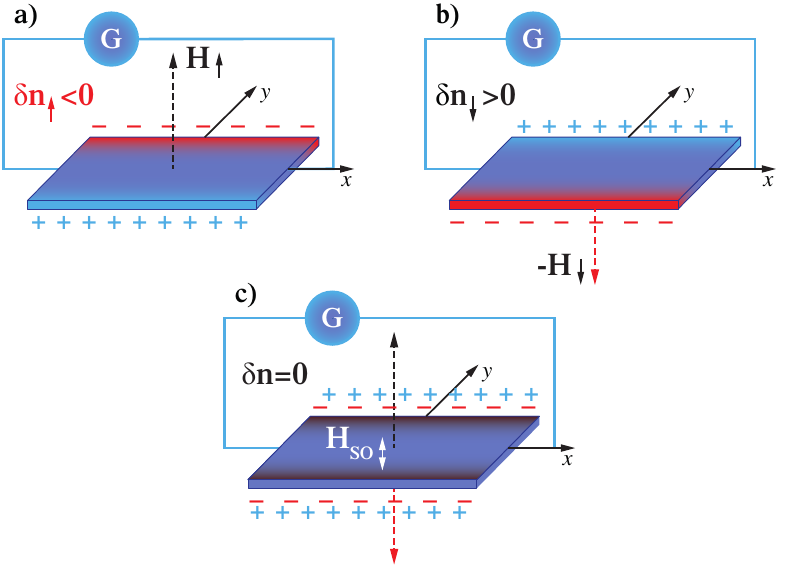}
   \end{tabular}
   \end{center}
   \caption[Fig2]
{ \label{fig:Fig2} : Schematic representation of the spin-Hall effect with the electrostatic charge accumulation $\delta n_{\updownarrow}$ at the boundaries. (a) usual Hall effect with the effective spin-orbit magnetic field $\vec H_{\uparrow}$ and (b) $\vec H_{\downarrow} = - \vec H_{\uparrow}$. (c ) the addition of  configurations (a) and (b) leads to an effective magnetic field acting on the two different electric carriers. }
   \end{figure}
The system under interest is a Hall bar of finite width contacted to an electric generator (see Fig.1), in which the invariance along the $x$ axis is assumed (the role of Corbino geometries \cite{Benda} or the presence of lateral contacts \cite{Popovic} are not under consideration here). The density of electric carriers is described along the $y$ direction, inside each spin-channels, as $n_{\updownarrow} = n_0 + \delta n_{\updownarrow}$, where $n_0$ is the density of electric carriers for an electrically neutral system, and $\delta n_{\updownarrow}$ is the accumulation of the electric charge along the $y$ axis. The charge accumulation is governed by the Poisson law, that defines the electric potential along the $y$ axis : $\nabla^2 V = \frac{\partial^2 V}{\partial y^2} =  - \frac{q \delta n}{\epsilon} $, where $\delta n = \delta n_{\uparrow} + \delta n_{\downarrow} $, $q$ is the electric charge, and $\epsilon$ is the electric permittivity. We assume that there is no accumulation of electric charges along the $x$ axis, so that the electric field is reduced to the drift force $n_0E_x^0$ in this direction, and $\partial n_{\updownarrow}/\partial x= 0$. 

The electro-chemical potential $\mu_{\updownarrow}$ is spin-dependent as it takes into account not only the electric potential $V$, but also the diffusion of the electric carriers due to the charge accumulation $\delta n_{\updownarrow}$, and the chemical potential $\mu^{ch}_{\updownarrow}$ that accounts for the spin-flip relaxation of the internal spin degrees of freedom (which is analogous to a chemical reaction \cite{DeGroot}). We have $\mu_{\updownarrow} = k \tilde T ln \left (\frac{n_{\updownarrow}}{n_0} \right ) + V + \mu^{ch}_{\updownarrow}$ \cite{Rubi,Moi2007,Entropy,MagDiff} where $k$ is the Boltzmann constant and the temperature $\tilde T$ is the Fermi temperature $\tilde T= T_F$ in the case of a fully degenerated conductors, or the temperature of the heat bath $\tilde T=T$ in the case of a non-degenerated semiconductors. 

The Ohm's law applied to the two channels reads $\vec J_{\updownarrow} = - q \hat \eta n_{\updownarrow} \vec \nabla \mu_{\updownarrow}$, where the mobility tensors $\hat \eta$ is a four by four matrix defined by the diagonal coefficients $\eta$ (the mobility of the charge carriers), and by the off-diagonal coefficients $\eta_{so}$ (the effective Hall mobility due to spin-orbit coupling). The off-diagonal coefficients obey the Onsager reciprocity relations $\eta_{xy \uparrow} = - \eta_{yx \uparrow} = \eta_{so}$ for the up spin channel and $\eta_{xy \downarrow} = - \eta_{yx \downarrow} = - \eta_{so}$ for the down spin-channel. The Ohm's law then reads  \cite{JPhys,EPL}:
\begin{equation}
\vec J_{\updownarrow}= q \eta n_{\updownarrow} \vec E_{\updownarrow} - D \vec \nabla n_{\updownarrow} \pm \vec e_z \times  \left ( - q n_{\updownarrow} \eta_{so} \vec E_{\updownarrow} +  D_{so} \vec \nabla n_{\updownarrow})\right )
\label{DP}
\end{equation}
where $\vec E_{\updownarrow} = - \vec \nabla (V + \mu_{\updownarrow}^{ch})$ and $D = \eta kT$, $D_{so} = \eta_{so} kT$ are the diffusion constants \cite{JPhys}. The DP equations are recovered in the case $\vec \nabla \mu_{\updownarrow}^{ch} = 0$.
The heat dissipation is due to the Joule heating for the two channels $- \vec J_{\updownarrow}. \vec \nabla \mu_{\updownarrow}$ and to the contribution due to the spin-flip relaxation. This last contribution reads  $\mathcal L \Delta \mu^2 $, where $\Delta \mu = \mu_{\uparrow} - \mu_{\downarrow}$ is the spin-accumulation potential, and $\mathcal L$ is the Onsager transport coefficient related to the spin-flip relaxation process \cite{DeGroot,PRB2000,Moi2007,Entropy}. Inserting equations (\ref{DP}) we have: 
\begin{equation}
P_J = \int_{\mathcal D} \left \{ q \eta n_{\uparrow} \left( \vec \nabla \mu_{\uparrow} \right)^2 + q \eta n_{\downarrow} \left( \vec \nabla \mu_{\downarrow} \right)^2 + \mathcal L \Delta \mu^2 \right \} {{d}^{3}}\vec{r}.
\label{PJ}
\end{equation}
where $\mathcal D$ is the volume of the device. Note that the expression $P_J$ of the Joule power is the same with and without Hall or Spin-Hall effects (i.e. with or without cross coefficients in the Ohm's law $\eta_{so}$, $D_{so}$ in Eq. (\ref{DP})), since these effects are nondissipative.

In order to illustrate the efficiency of the variational approach, we apply the Kirchhoff-Helmholz principle to the Joule power $P_J$ without any constraint : we observe that the functional derivative $\left ( \delta P_J/\delta \mu_{\updownarrow}\right )_{\mu_{\updownarrow}^{st}} = 0$ leads directly to the well known spin-accumulation equation that characterizes the stationary state for spin-injection through an interface between a ferromagnetic and a non-ferromagnetic conductor \cite{Johnson, Wyder,Valet-Fert, PRB2000,Shibata}:
\begin{equation}
\nabla^2 \Delta \mu^{st} - \frac{\Delta \mu^{st}}{l_{sf}^2}=0
\label{SpinAcc}
\end{equation} 
where $\Delta \mu^{st}$ is the stationary value for the spin-accumulation $\Delta \mu$ and the spin-diffusion length is given by the relation $1/l_{sf}^2 = 1/l_{\uparrow}^2 + 1/l_{\downarrow}^2$ with $ l_{\updownarrow} = \sqrt{q \eta n_{\updownarrow} /(4 \mathcal L)}$. 

However, the description of the lateral charge accumulation at the edges of the Hall bar imposes the electrostatic interactions (that takes the form of a Poisson's equation) to be introduced as a constraint with a first Lagrange multiplier $\lambda(y)$. Furthermore, in Hall devices, the electric generator imposes a constant current $J_x^{\circ}$ with the constant field $E_x^0$ along the $x$ axis, while the chemical potentials $\mu_{\updownarrow}(y)$ are left free along the $y$ axis. This constraint  is described by a second Lagrange multiplayer $\beta(y)$ related to the projection of equation (\ref{DP}) on the unit vector $\vec e_x$. The functional $\mathcal I$ to be optimized is given by:
\begin{eqnarray}
&&  \mathcal I  [\mu_{\updownarrow}, \vec \nabla \mu_{\updownarrow},  \nabla^2 \mu_{\updownarrow}, n_{\updownarrow},  \nabla^2 n_{\updownarrow}] = 
\int_{\mathcal D} \bigg \{ q \eta n_{\uparrow} \left(  \vec \nabla \mu_{\uparrow} \right)^2 + q \eta n_{\downarrow} \left(  \vec \nabla \mu_{\downarrow} \right)^2 + \mathcal L (\mu_{\uparrow} - \mu_{\downarrow})^2 \nonumber \\ 
&&  - \lambda(y) \left [ \nabla^2 \left( \mu_{\uparrow}+\mu_{\downarrow} \right) - \frac{k \tilde T}{q}  \nabla^2  \left (ln(n_{\uparrow}) + ln(n_{\downarrow}) \right ) +2q \frac{\delta n}{\epsilon} \right] \nonumber \\ 
&& - \beta(y) \left [ q\left (  \eta n_0 E_x^{\circ} \vec e_x -  \vec e_z \times \eta_{so}  \left ( n_{\uparrow} \vec \nabla \mu_{\uparrow} -  n_{\downarrow} \vec \nabla\mu_{\downarrow} \right) \right ). \vec e_x -  J_{x}^{\circ} \right] \, \, \bigg \}  \, dx dy.
\label{I_Min}
\end{eqnarray}
Note that this variational problem has been solved in a previous work \cite{Entropy} in the absence of spin-flip relaxation and with constant conductivity. Here, the minimization of Eq.(\ref{I_Min}) leads to the four Euler-Lagrange equations for the problem under interest. For simplicity, we will omit in what follows the superscript $st$ for the stationary values of the variables ($\mu_{\updownarrow} \equiv \mu_{\updownarrow}^{st} $, $\Delta \mu = \Delta \mu^{st}$, etc).
Thus, on one hand, the Euler-Lagrange equations $\frac{\delta \mathcal I}{\delta \mu_{\updownarrow}} = 0$ explictly read : 
\begin{eqnarray}
2 \mathcal L (\mu_{\uparrow} - \mu_{\downarrow}) - 2 \eta q \frac{\partial }{\partial y}\left (n_{\updownarrow} \frac{\partial \mu_{\updownarrow}}{\partial y} \right )  - \frac{\partial^2 \lambda}{\partial y^2} \mp q \eta_{so} \frac{\partial \left (\beta n_{\updownarrow} \right)}{\partial y} =0.
\label{deltaImu}
\end{eqnarray}
On the other hand, the Euler-Lagrange equations $\frac{\delta \mathcal I}{\delta n_{\updownarrow}} = 0$ are :
\begin{eqnarray}
 \eta q n_{\updownarrow} \left ( \frac{\partial \mu_{\updownarrow}}{\partial y} \right)^2 + \frac{k \tilde T}{q} \frac{\partial^2 \lambda}{\partial y^2} - \frac{2q n_{\updownarrow}}{\epsilon} \lambda \pm q \eta_{so} n_{\updownarrow}  \frac{\partial \mu_{\updownarrow}}{\partial y} \beta =0
\label{deltaIn}
\end{eqnarray}
On combining Eqs.(\ref{deltaIn}) into Eqs.(\ref{deltaImu}) we arrive at the relation between the Lagrange multipliers $\lambda$ and $\beta$:
\begin{eqnarray}
\lambda = - \frac{\epsilon kT}{2q^2} G_{\updownarrow}  + \frac{\epsilon}{2} \left [\eta \left ( \frac{\partial \mu_{\updownarrow}}{\partial y} \right )^2 \pm \beta \eta_{so} \frac{\partial \mu_{\updownarrow}}{\partial y}
  \right ],
\label{lambda}
\end{eqnarray}
where $G_{\updownarrow} = \frac{1}{n_{\updownarrow}}\left (  2 \eta q  \frac{\partial }{\partial y} \left ( n_{\updownarrow} \frac{\partial \mu_{\updownarrow}}{\partial y} \right ) \pm q \eta_{so} \frac{\partial }{\partial y}(\beta n_{\updownarrow}) \mp 2 \mathcal L \Delta \mu  \right )$. Injecting Eq.(\ref{lambda}) into Eq.(\ref{deltaImu}) yields:

\begin{eqnarray}
 \frac{\partial^2 G_{\updownarrow}}{\partial y^2} - \frac{2q^2 n_{\updownarrow}}{\epsilon kT} G_{\updownarrow} = \frac{q^2}{kT} \frac{\partial^2}{\partial y^2} \left [  \left (\eta  \frac{\partial \mu_{\updownarrow}}{\partial y} \right )^2  \pm  \beta \eta_{so}  \frac{\partial \mu_{\updownarrow}}{\partial y}  \right ].
\label{G}
\end{eqnarray}

The stationary state is verified if the Lagrange multipliers $\lambda$ and $\beta$ are related by Eqs.(\ref{lambda}) and (\ref{G}). The parameter $\beta$ is then free, provided Eq. (\ref{G}) is verified. Therefore, we choose $\beta$ such that the right hand side of Eq.(\ref{G}) vanishes, namely :

\begin{equation}
\beta = \mp \frac{\eta}{\eta_{so}}  \frac{\partial \mu_{\updownarrow}}{\partial y}
\label{beta}
\end{equation}
As $\beta$ does not depend on the spin state, it follows immediately that $\frac{\partial \mu_{\uparrow}}{\partial y} = - \frac{\partial \mu_{\downarrow}}{\partial y}$. Hence Eq.(\ref{G}) reduces to:

\begin{eqnarray}
\frac{\partial^2}{\partial y^2} \left ( \frac{n_0}{n_{\updownarrow}}  \frac{\partial }{\partial y} \left 
( \frac{n_{\updownarrow}}{n_0} \frac{\partial \mu_{\updownarrow}}{\partial y}  \right ) \mp \frac{\Delta \mu}{2 l_{sf}^2} \right ) - \frac{1}{\lambda_D^2} \left (  \frac{\partial }{\partial y} \left 
( \frac{n_{\updownarrow}}{n_0} \frac{\partial \mu_{\updownarrow}}{\partial y} \right ) \mp \frac{ \Delta \mu}{2 l_{sf}^2} \right ) = 0
\label{Result}
\end{eqnarray}

where we have introduced the Debye-Fermi length $\lambda_D = \sqrt{\frac{\epsilon k \tilde T}{2q^2 n_0}}$ and the spin-flip diffusion length $l_{sf} = \sqrt{\frac{q \eta n_0}{4 \mathcal L}}$ . The non-linear equation (\ref{Result}) is a fourth order differential equation for the chemical potential $\mu_{\updownarrow}$, that has no simple analytical solution. This equation together with the symmetry of the spin-dependent electric fields $\frac{\partial \mu_{\uparrow}}{\partial y} = - \frac{\partial \mu_{\downarrow}}{\partial y}$, is an {\it exact formulation of the stationary problem for the SHE}. This is the main result of this work. Interestingly, Eq.(\ref{Result}) does not depend on the Hall or Spin-Hall terms $\eta_{so}$ (as for the power $P_J$ in Eq.(\ref{PJ})), so that it can also be apply to the case of the diffusive spin-accumulation in the so-called non-local or lateral geometry. Note also that Eq.(\ref{Result}) is  non-trivial in the case without spin-flip relaxation $\Delta \mu = 0$, as discussed in reference  \cite{EPL}.

The physical significance of this result can be analyzed further by formulating Eq.(\ref{Result}) in terms of the current divergence $\vec{\nabla}\cdot\vec J_{\updownarrow}$. Inserting the divergence of Eqs.(\ref{DP}) into Eq.(\ref{Result}) yields:
\begin{eqnarray}
\frac{\partial^2}{\partial y^2}\left \{ \frac{n_0}{n_{\updownarrow}} \left ( \vec{\nabla}\cdot\vec J_{\updownarrow} \pm 2 \mathcal L \, \Delta \mu \right ) \right \}
- \frac{1}{\lambda_D^2} \left ( \vec{\nabla}\cdot\vec J_{\updownarrow} \pm 2 \mathcal L \, \Delta \mu \right ) = 0
\label{Conservation1}
\end{eqnarray}
A trivial solution is found with the usual conservation equations for the two channels

 $\vec{\nabla}\cdot\vec J_{\updownarrow} \pm 2 \mathcal L \, \Delta \mu = 0$. But the interesting point is that this simple solution corresponds to the situation in which the charge accumulation is ignored, which is the situation treated in the literature so far  \cite{Valet-Fert, PRB2000,Shibata, Zhang,Tse,Maekawa,Review, Saslow,JPhys}. In contrast, Eq.(\ref{Conservation1}) shows that due to electrostatic interactions surface currents are flowing within the region defined by the characteristic length $\lambda_D$.

In order to analyze further the solutions of Eq.(\ref{Result}), we assume a small charge accumulation $\delta n_{\updownarrow}/n_0 \ll1$ and use perturbation theory in terms of this small parameter. At zero order of perturbation we have:
\begin{eqnarray}
\frac{\partial^2}{\partial y} \left ( \frac{\partial^2 \mu_{\updownarrow}}{\partial y^2}  \mp \frac{\Delta \mu}{2l_{sf}^2} \right ) - \frac{1}{\lambda_D^2} \left (  \frac{\partial^2 \mu_{\updownarrow}}{\partial y^2}  \mp \frac{\Delta \mu}{2l_{sf}^2} \right ) = 0.
\label{ResultZero}
\end{eqnarray}
Due to the two characteristic length scales $\lambda_D$ and $l_{sf}$ (such that $\lambda_D/l_{sf} \ll 1$), we have to define the dimensionless variable  $\tilde y = y/\lambda_D$ (the limit $\lambda_D \rightarrow 0$ directly applied on Eq.(\ref{ResultZero}) is not correct due to the other limit $\lambda_D/l_{sf} \ll 1$). Taking the difference between the two channels in Eq.(\ref{ResultZero}) yields :
\begin{eqnarray}
\frac{\partial^4  \Delta \mu}{\partial \tilde y^4} - \left (  1+ \frac{\lambda_D^2}{l_{sf}^2} \right)  \frac{\partial^2 \Delta \mu}{\partial \tilde y^2}  + \frac{\lambda_D^2}{l_{sf}^2}\Delta \mu = 0
\label{ChargeSpinDiff0}
\end{eqnarray}

The limit $\lambda_D/l_{sf} \ll 1$ leads to $\frac{\partial^4  \Delta \mu}{\partial \tilde y^4} - \frac{\partial^2 \Delta \mu}{\partial \tilde y^2} = 0$, or, in terms of the variable $y$: 
\begin{equation}
\frac{{{\partial }^{4}}\Delta \mu }{\partial {{y}^{4}}}-\frac{1}{\lambda _{D}^{2}}\frac{{{\partial }^{2}}\Delta \mu }{\partial {{y}^{2}}} \approx 0
\label{SpinDiffSHE}
\end{equation}

Note that Eq.(\ref{SpinDiffSHE}) deviates from the well-known spin-accumulation equation Eq.(\ref{SpinAcc}) derived in the case of spin-injection. In particular, far away from the edges, we have:
\begin{equation}
\frac{{{\partial }^{2}}\Delta \mu }{\partial {{y}^{2}}}\approx 0,
\label{linear}
\end{equation}
hence the profile of the spin-accumulation $\Delta\mu(y)$ is linear in the bulk (i.e. for $y \gg \lambda_D$)  \cite{Bottegoni}. Inserting the solution $\frac{\partial \Delta \mu }{\partial y} = cst$ in the transport equation, we have 
\begin{equation}
\vec{J}_{\updownarrow} \cdot \vec{e}_{y}=0
\label{zeroCurrent}
\end{equation}
This stationary state is defined by zero spin-current, and an effective electric field such that $E_{\uparrow} = - E_{\downarrow}$. Analysis of the first order of perturbation of Eq.(\ref{Result}) shows that charges accumulate on the boundaries and therefore that the above discussion is unchanged.

Interestingly, the linear solution was also that found in the case without spin-flip scattering (i.e. with $l_{sf} \rightarrow \infty$) \cite{Benda}. This is due to the fact that, in the framework of the SHE, the spin-flip scattering is related to the free variables $\mu_{\updownarrow}$ or $J_{y \updownarrow}$. In other terms, the spin-flip relaxation process cannot force the Spin-Hall device to dissipate more at stationary state. This is the opposite in the case of the usual spin-injection that leads to the giant magnetoresistance effect, for which the spin-flip relaxation is related to the constrained variables $E_{x}^0$ or $J_{x \updownarrow}$: spin-flip scattering is then forced by the generator along the $x$ direction. 


At last, the physical meaning of the linear solution just found can be best understood by inserting Eq.(\ref{linear}) into the exact equation (\ref{Result}). We obtain the well-known screening equation for $\lambda_D/l_{sf} \ll 1$ and at first order in $\delta n / n_0$ :
\begin{equation}
\frac{{{\partial }^{2}} \delta n_{\updownarrow}}{\partial {{y}^{2}}} - \frac{\delta n_{\updownarrow}}{\lambda_D^2} = 0,
\label{Screening}
\end{equation}
Accordingly, this linear solution of the exact equation Eq.(\ref{Result}) is the stationary state that corresponds to {\it equilibrium} along the $y$ axis.

In conclusion, we studied the stationary state of the spin-Hall effect with taking into account both the electrostatic interactions and the spin-flip relaxation.  We defined the Spin-Hall effect by the corresponding DP transport equations and by the expression of the power. In the framework of the Kirchhoff-Helmoltz variational principle, the stationary state is defined by the minimization of the dissipated power under the constraints specified by the electrostatic interactions and by a uniform charge current injected along the $x$ axis. The minimization leads to a fourth order differential equation, that describes the system, including both surface and bulk currents and fields. This equation shows that the form of the usual conservation laws used in the context of spin-injection and giant magnetoresistance should be modified in order to take into account electrostatic interaction and screening effects.
We show that the solution for small charge accumulation and at the limit $\lambda_D/l_{sf} \ll 1$ is the same as that without spin-flip scattering, \textit{whatever the absolute value of} $l_{sf}$. This solution corresponds to the linear behavior of the spin-accumulation $\Delta \mu(y)$ and chemical potentials $\mu_{\updownarrow}(y)$. This analysis defines the ``effective electric fields'' $\partial \mu^{ch}_{\uparrow} / \partial y = - \partial \mu^{ch}_{\downarrow} / \partial y $, that compensates the ``effective Lorentz force'' generated by the spin-orbit scattering, and that leads to the observed spin-accumulation field $\partial \Delta \mu / \partial y$. 



\begin{thebibliography}{0}

\bibitem{Dyakonov} M. I. Dyakonov, and  V. I. Perel, {\it Possibility of orienting electron spins with current} ZhETF Pis. Red. {\bf 13}, no 11, 657 (1971) and and  {\it Current induced spin orientation in semiconductorsÊ}, Phys. Lett A 35, 459 (1971).
\bibitem{Dyakonov2} M. I. Dyakonov, {\it spin Physics in Semiconductors}, Springer Series in Solid-States Sciences 2008.
\bibitem{Hirsch} J. E. Hirsch {\it Spin Hall effect} Phys. Rev. Lett. {\bf 83}, 1834 (1999). 
\bibitem{Zhang} Sh. Zhang, {\it Spin Hall effect in the presence of spin diffusion}, Phys. Rev. Lett. {\bf 85}, 393 (2000).
\bibitem{Tse}  W.-K. Tse, J. Fabian I. $\check{Z}$uti\'c, and S. Das Sarma, {\it Spin accumulation in the extrinsic spin Hall effect}, Phys. Rev. B {\bf 72}, 241303(R) (2005)
\bibitem{Maekawa} S. Takahashi and S. Maekawa {\it Spin current, spin accumulation and spin Hall effect}, Sci. Technol. Adv. Mater. {\bf 9} (2008) 014105. 
\bibitem{Review} J. Sinova, S. O. Valenzuela, J. Wunderlich, C. H. Back, T. Jungwirth {\it Spin Hall effects}, Rev. Mod. Phys. {\bf 87}, 1213 (2015).
\bibitem{Saslow} W. M. Saslow, {\it Spin Hall effect and irreversible thermodynamics: center-to-edge transverse current-induced voltage} Phys. Rev. B {\bf 91}, 014401 (2015).
\bibitem{JPhys} J.-E. Wegrowe,{ \it Twofold stationary states in the classical spin-Hall effect} J. Phys.: Cond Matter {\bf 29}, 485801 (2017).
\bibitem{EPL} J.-E. Wegrowe, R. V. Benda, and J. M. Rubi, {\it Conditions for the generation of sin current in spin-Hall devices}, Europhys. Lett {\bf 18} 67005 (2017).
\bibitem{Awschalom} Y. K. Kato, R. C. Myers, A. C. Gossard, D. D. Awschalom, {\it Observation of the spin Hall effect in semiconductors}, Science {\bf 306} 1910 (2004). 
\bibitem{Jungwirth} J. Wunderlich; B. Kaestner; J. Sinova; T. Jungwirth, {\it Experimental Observation of the Spin-Hall Effect in a Two-Dimensional Spin-Orbit Coupled Semiconductor System}, Phys. Rev. Lett. {\bf 94}, 047204  (2005).
\bibitem{Valenzuela} S. O. Valenzuela and M. Tinkham {\it Direct electronic measurement of the spin-Hall effect}, Nature {\bf 442}, 176 (2006).
\bibitem{Otani} T. Kimura; Y. Otani; T. Sato; S. Takahashi; S. Maekawa {\it Room-Temperature Reversible Spin Hall Effect}, Phys. Rev. Lett. {\bf 98}, 156601 (2007).
\bibitem{Gambardella} C. Stamm, C. Murer, M. Berritta, J. Feng, M. Gaburac, P. M. Oppeneer, and P Gambardella, {\it Magneto-optical detection  of the Spin Hall effect in Pt and W thin films}, Phys. Rev. Lett. {\bf 119}, 087203 (2017).
\bibitem{Bottegoni} F. Bottegoni, C. Zucchetti, S. Dal Conte, J. Frigerio, E. Carpene, C. Vergnaud, M. Jamet, G. Isella, F. Ciccacci, G. Cerullo, and M. Finazzi, Phys. Rev. Lett. {\bf 118}, 167402 (2017).
\bibitem{Valet-Fert} T. Valet and A. Fert, {\it Theory of the perpendicular magnetoresistance in magnetic mutilayers}, Phys. Rev. B {\bf 48}, (1993)
\bibitem{PRB2000} J.-E. Wegrowe, {\it Thermokinetic approach of the generalized Landau-Lifshitz-Gilbert equation with spin-polarized current}, Phys. Rev. B {\bf 62}, (2000), 1067. 
\bibitem{Shibata} J. Shibata and H. Kohno, {\it Spin and charge transport induced by gauge fields in a ferromagnet}, Phys. Rev. B, {\bf 84}, 184408 (2011)
\bibitem{Jaynes} E. T. Jaynes, {\it The minimum entropy production principle}, Ann. Rev. Phys. Chem. {\bf 31} 579 - 601 (1980).
\bibitem{Benda} R. V. Benda, J. M. Rubi and J.-E. Wegrowe, {\it toward Joule heating optimization in Hall devices}, submitted (2018). 
\bibitem{Popovic} R.S. Popovic, {\it Hall Effect Devices}, IoP Publishing, Bristol and Philadelphia 2004.
\bibitem{DeGroot} See the sections {\it relaxation phenomena} and {\it internal degrees of freedom} in Chapter 10 of De Groot, S.R.; Mazur, P. {\it Non-equilibrium Thermodynamics}; North-Holland: Amsterdam, The Netherlands, 1962. .

\bibitem{Rubi} D. Reguera, J. M. G. Vilar, and J. M. Rub\`i. {\it Mesoscopic Dynamics of Thermodynamic Systems}, J. Phys. Chem. B 109 (2005).
\bibitem{Moi2007} J.-E. Wegrowe, M.-C. Ciornei, H.-J. Drouhin, {\it Spin transfer in an open ferromagnetic layer: from negative damping to effective temperature}, J. Phys.:Condens. Matter {\bf 19} 165213 (2007).
\bibitem{Entropy} J.-E. Wegrowe, H.-J. Drouhin, {\it Spin-currents and spin pumping forces for spintronics}, Entropy {\bf 13}, 316 (2011). 
\bibitem{MagDiff} J.-E. Wegrowe, S. N. Santos, M.-C. Ciornei, H.-J. Drouhin, J. M. Rub\'i, {\it Magnetization reversal driven by spin-injection: a diffusive spin-transfer effect}, Phys. Rev. B {\bf 77} , 174408 (2008).

\bibitem{Johnson} M. Johnson and R. H. Silsbee, {\it "Interfacial charge-spin coupling: injection and detection of spin magnetization in metal"}, Phys. Rev. Lett. {\bf 55}, 1790 (1985) 
\bibitem{Wyder} P.C. van son, H. van Kempen, and P. Wyder, {\it Boundary resistance of the ferromagnetic-nonferromagnetic metal interface}  Phys. Rev. Lett. {\bf 58}, 2271 (1987)


 


\end{thebibliography}
\end{document}